\documentstyle[times,pramana,epsf,floats]{ias}
\def\v{\begingroup\obeyspaces\u}
\def\u#1{\tt#1\endgroup}
\def\IQ{\v{IQ}}
\def\IL{\v{IL}}

\def\IN{\v{IN}}
\def\IC{\v{IC}}

\def\ISQ{\v{ISQ}}
 \def\ino{\widetilde}\def\gluino{\ino{g}}
 \def\squark{\ino{q}} %\def\sup{\ino{U}}\def\sdn{\ino{D}}
 \def\slepton{\ino{\ell}}  \def\snu{\ino{\nu}}
 \def\gaugino{\ino{\chi}} \def\ntlino#1{\ino{\chi}^0_{#1}}
\def\s#1{{\small#1}}

\def\TA{{\small TAUOLA}}
\def\HW{\s{HERWIG}}

\def\IS{\s{ISAJET}}
\def\IW{\s{ISAWIG}}

\def\MSSM{{MSSM}}
\def\SY{\s{SUSY}}
\def\QCD{\s{QCD}}

\def\qbar{\bar{q}}
\def\Qbar{\bar{Q}}
\def\dbar{\bar{d}}
\def\ubar{\bar{u}}

\def\B0bar{\overline{B^0}}

\def\l{\ell}
\begin{document}
\mark{{HERWIG: an event generator for MSSM processes}{Stefano Moretti}}
\title{HERWIG: an event generator for MSSM processes}

\author{Stefano Moretti}
\address{CERN Theory Division, CH-1211 Geneva 23, Switzerland
and\\
Institute for Particle Physics Phenomenology,
University of Durham, Durham DH1 3LE, UK}
\keywords{Monte Carlo programs, Event generation, Supersymmetry,
MSSM processes}
\pacs{24.10.Lx, 11.30.Pb, 12.60.Jv}
\abstract{
The \HW\ event generator was widely used throughout the workshop,
particularly in the emulation of Supersymmetric (\SY) and Higgs processes
in the context of the Minimal Supersymmetric Standard Model (\MSSM).
We briefly review here its main features in this respect.}

\maketitle

\vskip-7.85cm

\noindent
\hskip9.65cm{CERN-TH/2002-075}

\noindent
\hskip9.65cm{IPPP/02/22}

\noindent
\hskip9.65cm{DCPT/02/44}

\noindent
\hskip9.65cm{April 2002}

\vskip+7.85cm

\section{Introduction}

    \HW\ is a general-purpose Monte Carlo (MC) event generator for high-energy
    processes,  providing a full simulation of
    hard lepton-lepton, lepton-hadron and  hadron-hadron scattering
    and soft hadron-hadron collisions in a single package, comprising:
\begin{itemize}
\item
    Initial- and final-state \QCD\ jet evolution with soft gluon
    interference taken into account via angular ordering;
\item
    Colour coherence of (initial and final) partons in all hard
subprocesses, including the production and decay of heavy quarks and
\SY\  particles;
\item Spin correlations in the decay of heavy fermions;
\item Lepton beam polarisation in selected processes;
\item
    Azimuthal correlations  within and  between jets due to gluon
interference and polarisation;
\item
    A cluster model for jet hadronisation based on non-perturbative
gluon splitting, and a similar  cluster model for soft and underlying
hadronic  events;
\item
    A space-time picture of event development, from parton showers to
hadronic decays, with an optional colour rearrangement model based on
space-time structure.
\end{itemize}
Several of these features were already present in \HW\ versions 5.1--5.9 and
were described accordingly in some detail in Ref.~\cite{HERWIG59}.
The \HW\ source codes, together with other useful files and information,
can be obtained from the following web site:
\small\begin{quote}\tt
      {\small http://hepwww.rl.ac.uk/theory/seymour/herwig/.}
\end{quote}\normalsize

\section{The MSSM in HERWIG}

Starting from version 6.1 \cite{HERWIG61}, 
\HW\ includes the production and decay of (s)particles,
as given by the \MSSM. A detailed description of their
implementation can be found in \cite{SUSYWIG}, which complements
the current \HW\ manual \cite{HERWIG62}. The last two successive
versions are documented in \cite{HERWIG63-64}. 

The \HW\ particle content is listed in
Tab.~1. For sparticles that mix, the subscripts label
the mass eigenstates in the ascending order of mass.
The two Higgs Doublet Model (2HDM) Higgs sector, intrinsic to the
\MSSM, is also included. The three neutral Higgs bosons are denoted
by $h^0$, $H^0$ and $A^0$, whereas the charged ones by $H^\pm$.

%\begin{table}[h]
\begin{center}
%\label{tab:MSSM}
{\small {\bf Table 1.} The \MSSM\ particle spectrum implemented in \HW.}
\small
\begin{tabular}{|lcclcc|}     \hline
%&&&&&\\  
Particle & & Spin & Particle & & Spin \\
%&&&&& \\
\hline
&&&&&\\
quark  & $q$ & 1/2 & squarks & $\ino{q}_{L,R}$ & 0 \\
charged lepton & $\ell$ & 1/2 & charged sleptons & $\ino{\ell}_{L,R}$ & 0 
\\
neutrino  & $\nu$ & 1/2 &  sneutrino & $\ino{\nu}$ & 0 \\
gluon & $g$ & 1 & gluino & $\ino{g}$ & 1/2 \\
photon & $\gamma$ & 1 & photino & $\ino{\gamma}$ & 1/2 \\
neutral gauge boson &$Z^0$ & 1 & zino & $\ino{Z}$ & 1/2 \\
neutral Higgs bosons & $h^0,H^0,A^0$ & 0 & neutral Higgsinos &
$\ino{H}^{0}_{1,2}$ & 1/2 \\ 
charged gauge boson & $W^\pm$ & 1 & wino & $\ino{W}^\pm$ & 1/2 \\
charged Higgs boson & $H^\pm$ & 0 & charged Higgsino &
$\ino{H}^{\pm}$ & 1/2 \\ 
graviton & $G$ & 2 & gravitino & $\ino{G}$ & 3/2 \\
&&&&& \\ \hline
%&&&&& \\ 
\multicolumn{6}{|c|}{
$\ino{W}^\pm, \ino{H}^\pm$ mix to form 2 chargino 
mass eigenstates $\gaugino^{\pm}_{1}, \gaugino^{\pm}_{2}$}\\
\multicolumn{6}{|c|}{
$\ino{\gamma}$, $\ino{Z}$, $\ino{H}^0_{1,2}$ mix to
form 4
neutralino mass eigenstates $\ntlino{1},\ntlino{2},\ntlino{3},\ntlino{4}$}\\ 
\multicolumn{6}{|c|}{${\ino t}_L,{\ino t}_R$ (and similarly
${\ino b}, {\ino\tau}$) mix to form the mass eigenstates
${\ino t}_1, {\ino t}_2$} \\ 
%\multicolumn{6}{|c|}{~~~~~~~~~~~}\\
\hline
\end{tabular}
\end{center}
%\end{table}

\HW\ does not contain any  built-in models for Supersymmetry-breaking  
scenarios. In all cases the general \MSSM\ particle spectrum
and decay tables must be provided just like those for any other
particles. A package, \IW, has been created to work with \IS\ \cite{ISAJET}
to produce a file containing the \SY\ particle masses, lifetimes,
couplings and mixing 
parameters. This package takes  the outputs of the \IS\
\MSSM\ programs and produces a data file in a format that can be read into
\HW\ for the subsequent process generation. The user can
produce a similar file provided that the correct format is used.
To this end, we invite the consultation of the \IW\ webpage:
\small\begin{quote}\tt
{\small 
http://www-thphys.physics.ox.ac.uk/users/PeterRichardson/HERWIG/isawig.html}
\end{quote}\normalsize
where some example of input files can be found (the \SY\ benchmark 
points recommended in Ref.~\cite{SPS} are already available). Our
\MSSM\ conventions are described in \cite{SUSYWIG}.
The mass spectrum and decay modes, being read from input files,
are completely general, however, the following caveats are
to be borne in mind:
(i) \SY\ particles do not radiate (which is reasonable if their
decay lifetimes are much shorter than the QCD confinement scale); 
(ii) CP-violating \SY\ phases are not included.

In addition to the decay modes implemented in the \IS\ package \IW\ also
includes the calculation of all 2-body squark/slepton 
and 3-body gaugino/gluino R-parity violating (RPV) decay modes 
(alas, RPV lepton-gaugino and slepton-Higgs mixing is not considered).

The emulation of RPV processes is also a feature of the production
stage. Tab.~2 illustrates all \MSSM\ modes 
available at present (version 6.4). Here, {\tt IPROC} is the input
label selecting the hard process.

%\begin{table}[h]
\begin{center}
%\label{tab:hard}
{\small {\bf Table 2.} The \MSSM\ hard scattering processes implemented in \HW.}
\small
\begin{tabular}{|c|l|}
\hline
 {\tt IPROC} &        2$\to$ 2 \MSSM\ processes in $\ell^+\ell^-$ ($\ell=e,\mu$) \\
\hline
   700-99  & R-parity conserving \SY\ processes \\
   700     & $\l^+ \l^- \to$~2-sparticle processes (sum of 710--760)\\
   710     & $\l^+ \l^- \to$~neutralino pairs (all neutralinos) \\
706+4{\tt IN1}+{\tt IN2} &$\l^+ \l^- \to \gaugino^0_{\mbox{\scriptsize IN1}}
                         \gaugino^0_{\mbox{\scriptsize IN2}}$
                      ({\tt IN1,2}=neutralino mass eigenstate)\\
   730     & $\l^+ \l^- \to$~chargino pairs (all charginos) \\
728+2{\tt IC1}+{\tt IC2} &$\l^+ \l^- \to \gaugino^+_{\mbox{\scriptsize IC1}}
                         \gaugino^-_{\mbox{\scriptsize IC2}}$
                      ({\tt IC1,2}=chargino mass eigenstate) \\
   740     & $\l^+ \l^- \to$~slepton pairs (all flavours) \\
   736+5\IL& $\l^+ \l^- \to \slepton_{L,R} \slepton_{L,R}^*$
             ($\IL=1,2,3$ for $\slepton=\tilde{e},\tilde{\mu},\tilde{\tau}$) \\
   737+5\IL& $\l^+ \l^- \to \slepton_{L} \slepton_{L}^*$ (\IL\ as above) \\
   738+5\IL& $\l^+ \l^- \to \slepton_{L} \slepton_{R}^*$ (\IL\ as above)\\
   739+5\IL& $\l^+ \l^- \to \slepton_{R} \slepton_{R}^*$ (\IL\ as above)\\
   740+5\IL& $\l^+ \l^- \to \snu_{L} \snu_{L}^*$ 
             ($\IL=1,2,3$ for $\snu_e, \snu_\mu, \snu_\tau$) \\
   760      & $\l^+ \l^- \to$~squark pairs (all flavours) \\
   757+4\IQ & $\l^+ \l^- \to \squark_{L,R} \squark^*_{L,R}$
             ($\IQ=1...6$ for $\squark=\tilde{d}...\tilde{t}$)\\
   758+4\IQ & $\l^+ \l^- \to \squark_{L} \squark^*_{L}$
                (\IQ\ as above)\\
   759+4\IQ & $\l^+ \l^- \to \squark_{L} \squark^*_{R}$
                (\IQ\ as above)\\
   760+4\IQ & $\l^+ \l^- \to \squark_{R} \squark^*_{R}$
                (\IQ\ as above)\\
\hline
\end{tabular}
\end{center}
%\end{table}

%\begin{table}[h]
\begin{center}
{\small {\bf Table~2.} Continues.}\\
\small
\begin{tabular}{|c|l|}
\hline
 {\tt IPROC} &  2$\to$ 2 \MSSM\ processes in $\ell^+\ell^-$ ($\ell=e,\mu$)               \\
\hline
   800-99  & R-parity violating \SY\ processes \\
   800     & Single sparticle production, sum of 810--840 \\
   810     & $\l^+ \l^- \to \gaugino^0 \nu_i$, (all neutralinos)\\
   810+\IN & $\l^+ \l^- \to \gaugino^0_{\mbox{\scriptsize IN}} \nu_i$,
             (\IN=neutralino mass state)\\
   820     & $\l^+ \l^- \to \gaugino^- e^+_i$ (all charginos) \\
   820+\IC & $\l^+ \l^- \to \gaugino^-_{\mbox{\scriptsize IC}} e^+_i$,
             (\IC=chargino mass state) \\
   830     & $\l^+ \l^- \to \snu_i Z^0$ and 
             $\l^+ \l^- \to \slepton^+_i W^-$  \\
   840     & $\l^+ \l^- \to \snu_i h^0/H^0/A^0$ and 
             $\l^+ \l^- \to \slepton^+_i H^-$  \\
   850     & $\l^+ \l^- \to \snu_i \gamma$ \\
   860     & Sum of 870 and 880 \\
   870     & $\l^+ \l^- \to \l^+ \l^-$, via LLE only \\
   867+3{\tt IL1}+{\tt IL2} & 
$\l^+ \l^- \to \l^+_{\mbox{\scriptsize IL1}} \l^-_{\mbox{\scriptsize IL2}}$
          ({\tt IL1,2}=1,2,3 for $e,\mu,\tau$) \\
   880     & $\l^+ \l^- \to \bar d  d$, via LLE and LQD \\
   877+3{\tt IQ1}+{\tt IQ2} & 
$\l^+ \l^- \to d_{\mbox{\scriptsize IL1}} \bar d_{\mbox{\scriptsize IL2}}$
          ({\tt IQ1,2}=1,2,3 for $d,s,b$) \\
\hline
\end{tabular}
\end{center}
%\end{table}

%\begin{table}[h]
\begin{center}
{\small {\bf Table~2.} Continues.}\\
\small
\begin{tabular}{|c|l|}
\hline
 {\tt IPROC} &  2$\to$ 2 \MSSM\ processes in hadron-hadron               \\
\hline
   3000-999& R-parity conserving \SY\ processes\\
   3000    & 2-parton $\to$ 2-sparticle processes (sum of those below)\\
   3010    & 2-parton $\to$ 2-sparton processes \\
   3020    & 2-parton $\to$ 2-gaugino processes \\
   3030    & 2-parton $\to$ 2-slepton processes \\
\hline
\end{tabular}
\end{center}
%\end{table}
\vskip1.0cm
%\begin{table}[h]
\begin{center}
{\small {\bf Table~2.} Continues.}\\
\small
\begin{tabular}{|c|l|}
\hline
 \v{IPROC} &     2$\to$ 2 \MSSM\ processes in hadron-hadron     \\
\hline
   4000-99   &  R-parity violating \SY\ processes via LQD\\
   4000      & single sparticle production, sum of 4010--4050 \\
   4010      & $\ubar_j d_k \to \gaugino^0 l^-_i$,
               $\dbar_j d_k \to \gaugino^0 \nu_i$ (all neutralinos)\\
  4010+\IN  & $\ubar_j d_k \to \gaugino^0_{\mbox{\scriptsize IN}} l^-_i$,
               $\dbar_j d_k \to \gaugino^0_{\mbox{\scriptsize IN}} \nu_i$
(\IN=neutralino mass state)\\
   4020      & $\ubar_j d_k \to \gaugino^- \nu_i$, 
               $\dbar_j d_k \to \gaugino^- e^+_i$ (all charginos) \\
   4020+\IC & $\ubar_j d_k \to \gaugino^-_{\mbox{\scriptsize IC}} \nu_i$,
               $\dbar_j d_k \to \gaugino^-_{\mbox{\scriptsize IC}} e^+_i$ (\IC=chargino mass state) \\
   4040      & $u_j \dbar_k \to \tilde{\tau}^+_i Z^0$,
               $u_j \dbar_k \to \snu_i W^+$ and 
               $d_j \dbar_k \to \slepton^+_i W^-$  \\
   4050      & $u_j \dbar_k \to \slepton^+_i h^0/H^0/A^0$,
               $u_j \dbar_k \to \snu_i H^+$ and 
               $d_j \dbar_k \to \slepton^+_i H^-$  \\
   4060      & Sum of 4070 and 4080 \\
   4070      & $\ubar_j d_k \to \ubar_l d_m $ and 
               $\dbar_j d_k \to \dbar_l d_m $, via LQD only \\
   4080      & $\ubar_j d_k \to \nu_j l^-_k $ and 
               $\dbar_j d_k \to l^+_j l^-_k $, via LQD and LLE \\
\hline
   4100-99 & R-parity violating \SY\ processes via UDD\\
   4100    & single sparticle production, sum of 4110--4150\\
   4110      & $u_i d_j \to \gaugino^0 \dbar_k$,  
               $d_j d_k \to \gaugino^0 \bar{u_i}$ (all neutralinos)\\
  4110 +\IN  & $u_i d_j \to \gaugino^0_{\mbox{\scriptsize IN}} \dbar_k$,  
               $d_j d_k \to \gaugino^0_{\mbox{\scriptsize IN}}
 \bar{u_i}$(\IN\ as above)\\
  4120       & $u_i d_j \to \gaugino^+ \ubar_k$,  
               $d_j d_k \to \gaugino^- \bar{d_i}$  (all charginos) \\
  4120 +\IC  & $u_i d_j \to \gaugino^+_{\mbox{\scriptsize IC}} \ubar_k$,  
               $d_j d_k \to \gaugino^-_{\mbox{\scriptsize IC}}
 \bar{d_i}$  (\IC\ as above) \\
  4130       & $u_i d_j \to \gluino \dbar_k$,  
               $d_j d_k \to \gluino \bar{u_i}$ \\
  4140       & $u_i d_j \to \tilde{b}^*_1 Z^0$,
$d_j d_k \to \tilde{t}^*_1 Z^0$, 
               $u_i d_j \to \tilde{t}^*_i W^+$
and $d_j d_k \to \tilde{b}^*_i W^-$  \\
  4150       & $u_i d_j \to \tilde{d}^*_{k1} h^0/H^0/A^0$, 
               $d_j d_k \to \tilde{u}^*_{i1} h^0/H^0/A^0$,
               $u_i d_j \to \tilde{u}^*_{k\alpha} H^+$, 
               $d_j d_k \to \tilde{d}^*_{i\alpha} H^-$  \\
  4160       & $u_i d_j \to u_l d_m$, $d_j d_k \to d_l d_m$ via UDD. \\
\hline
\end{tabular}
\end{center}
%\end{table}

%\begin{table}[h]
\begin{center}
{\small {\bf Table~2.} Continues.}\\
\small
\begin{tabular}{|c|l|}
\hline
 {\tt IPROC} &   $2\to2$ \MSSM\ Higgs processes in $\ell^+\ell^-$ ($\ell=e,\mu$)  \\
\hline
    910    &      $\ell^+ \ell^- \to \nu_e \bar\nu_e h^0 + e^+ e^- h^0$\\
    920    &      $\ell^+ \ell^- \to \nu_e \bar\nu_e H^0 + e^+ e^- H^0$\\
\hline
    960    &      $\ell^+ \ell^- \to Z^0 h^0$\\ 
    970    &      $\ell^+ \ell^- \to Z^0 H^0$\\ 
\hline
    955    &      $\ell^+ \ell^- \to H^+ H^-$\\
    965    &      $\ell^+ \ell^- \to A^0 h^0$\\
    965    &      $\ell^+ \ell^- \to A^0 H^0$\\
\hline
\end{tabular}
\end{center}
%\end{table}

%\begin{table}[h]
\begin{center}
{\small {\bf Table~2.} Continues.}\\
\small
\begin{tabular}{|c|l|}
\hline
 {\tt IPROC} &   $2\to2$ \MSSM\ Higgs processes in hadron-hadron  \\
\hline
   3310,3315    & $q\qbar' \to W^\pm h^0,H^\pm h^0$  \\
   3320,3325    & $q\qbar' \to W^\pm H^0,H^\pm H^0$  \\
      3335      & $q\qbar' \to H^\pm A^0$ \\
      3350      & $q\qbar  \to W^\pm H^\mp$  \\
      3355      & $q\qbar  \to H^\pm H^\mp $  \\
   3360,3365    & $q\qbar  \to Z^0 h^0,A^0 h^0$ \\
   3370,3375    & $q\qbar  \to Z^0 H^0,A^0 H^0$ \\
\hline
   3410      & $bg \to b~h^0$ + ch.\ conj.\\
   3420      & $bg \to b~H^0$ + ch.\ conj.\\
   3430      & $bg \to b~A^0$ + ch.\ conj.\\
   3450      & $bg \to t~H^-$ + ch.\ conj.\\
\hline
   3610& $q\qbar/gg \to h^0$  \\
   3620& $q\qbar/gg \to H^0$  \\
   3630& $q\qbar/gg \to A^0$ \\
\hline
\end{tabular}
\end{center}
%\end{table}

%\begin{table}[h]
\begin{center}
{\small {\bf Table~2.} Continues. (For the definition of \ISQ, see
\cite{HERWIG63-64}.)}\\
\small
\begin{tabular}{|c|l|}
\hline
 {\tt IPROC} &   $2\to3$ \MSSM\ Higgs processes in hadron-hadron                    \\
\hline
   3100+\ISQ& $gg/q\qbar\to {\tilde q}{\tilde q}^{'*} {H^\pm}$ \\
%   (see first Table below) \\
\hline
   3200+\ISQ& $gg/q\qbar\to {\tilde q}{\tilde q}^{'*} {h,H,A}$ \\
%   (see second Table below) \\
\hline
   3500     & $b q \to b q' H^\pm$ + ch.\ conj. \\
\hline
   3710& $q\qbar \to q'\qbar' h^0$  \\
   3720& $q\qbar \to q'\qbar' H^0$  \\
\hline
   3810+\IQ& $gg+q\qbar\to Q\Qbar h^0$ (\IQ\ for $Q$ flavour) \\
   3820+\IQ& $gg+q\qbar\to Q\Qbar H^0$ ('') \\
   3830+\IQ& $gg+q\qbar\to Q\Qbar A^0$ ('') \\
3839~~~~~~~& $gg+q\qbar\to b\bar t H^+$ + ch. conjg. \\
   3840+\IQ& $gg       \to Q\Qbar h^0$ (\IQ\ as above) \\
   3850+\IQ& $gg       \to Q\Qbar H^0$ ('') \\
   3860+\IQ& $gg       \to Q\Qbar A^0$ ('') \\
3869~~~~~~~& $gg       \to b\bar t H^+$ + ch. conjg.  \\
   3870+\IQ& $q\qbar   \to Q\Qbar h^0$ (\IQ\ as above) \\
   3880+\IQ& $q\qbar   \to Q\Qbar H^0$ ('') \\
   3890+\IQ& $q\qbar   \to Q\Qbar A^0$ ('') \\
3899~~~~~~~& $q\qbar   \to b\bar t H^+$ + ch. conjg. \\
\hline
\end{tabular}
\end{center}
%\end{table}

The Matrix Elements (MEs) used for the hard scatterings have been listed in
\cite{SUSYWIG}, with the treatment of the different colour connections
in \SY\ QCD processes described in \cite{Colour} (which also superseeds
the standard QCD algorithm of \cite{HERWIG59}). 

Spin  correlations  are available in  processes where 
\SY\ particles are produced (and top quarks and $\tau$ leptons
as well), as described in \cite{Spin}. Whenever these particles
decay,  3- and 4-body MEs are used to describe their
dynamics (with or without the spin correlations). In the case 
of $\tau$ leptons, an interface to \TA\ \cite{TAUOLA} is also available.

Finally, polarisation has been implemented for incoming leptonic beams in
\SY\ processes. These effects are included both
    in the  production  of \SY\  particles  and via  the  spin  correlation
    algorithm in their decays.

%\section*{Acknowledgements} I thank the organisers of the workshop
%for the excellent atmosphere and stimulating environment that they 
%have created during the workshop.

\end{document}